\documentclass[aps,preprint,amsmath,amssymb]{revtex4}
\usepackage{psfig}
\usepackage{graphicx}

\newcommand{\nn}{\nonumber}
\newcommand{\bd}{\begin{document}}
\newcommand{\ed}{\end{document}}
\newcommand{\bc}{\begin{center}}
\newcommand{\ec}{\end{center}}
\newcommand{\be}{\begin{eqnarray}}
\newcommand{\ee}{\end{eqnarray}}
\renewcommand{\thefootnote}{\alph{footnote}}
\newcommand{\se}{\section}
\newcommand{\sse}{\subsection}
\newcommand{\bi}{\bibitem}
\def\figcap{\section*{Figure Captions\markboth
     {FIGURECAPTIONS}{FIGURECAPTIONS}}\list
     {Figure \arabic{enumi}:\hfill}{\settowidth\labelwidth{Figure 999:}
     \leftmargin\labelwidth
     \advance\leftmargin\labelsep\usecounter{enumi}}}
\let\endfigcap\endlist \relax

\begin{document}

\title{\large \bf Forward-backward Asymmetry in $K^+\to\pi^+ \ell^+\ell^-$ }

\author{ \bf Chuan-Hung Chen$^{a}$, C.~Q.~Geng$^{b,c}$ and I-Lin Ho$^{b}$}
 \affiliation{  $^{a}$Institute of
Physics, Academia Sinica, Taipei, Taiwan 115, Republic of China \\
$^{b}$Department of Physics, National Tsing Hua
University, Hsinchu, Taiwan 300, Republic of China \\
$^{c}$ Theory Group, TRIUMF, 4004 Wesbrook Mall, Vancouver, B.C.
V6T 2A3, Canada }

\date{\today}

\begin{abstract}
We study the forward-backward asymmetries in the decays of
$K^+\to\pi^+\ell^+ \ell^-$ ($\ell=e$ and $\mu$) in the presence of
scalar or tensor terms. We find that with the scalar (tensor) type
interaction the asymmetry can be up to ${\cal O}(10^{-3})$ (${\cal
O}(10^{-1}))$  and arbitrary large for the electron and muon
modes, respectively, without conflict with the experimental data.
We also discuss the cases in the minimal supersymmetric standard
model where the scalar terms can be induced. In particular we show
that the asymmetry in $K^+\to\pi^+\mu^+\mu^-$ can be as large as
${\cal O}(10^{-3})$ in the large $\tan\beta$ limit, which can be
tested in future experiments such as CKM at Fermilab.
\end{abstract}

 \maketitle

\section{Introduction}

The flavor-changing neutral current (FCNC) processes of $K^{\pm}\to
\pi^{\pm} \ell^+ \ell^-\ (\ell=e,\mu)$ are suppressed
and dominated by the long distance (LD) contributions involving one
photon exchange \cite{k1,k2,k3,k4} in the standard model (SM).
The decays have been successfully described within the framework of
chiral perturbation theory (ChPT) \cite{ChPT0} including electroweak
interactions at
${\cal O}(p^6)$ \cite{ChPT1}
in terms of a vector
interaction form factor fixed by experiments.
However, it is important to compare the measurements in the two decays to
see if there are differences in the form factors since
they would indicate new physics.
Recently, the vector form factor has been determined by the
 high precision measurement on the electron mode by
the BNL-E865 Collaboration
\cite{E865e} at the
Brookhaven Alternating Gradient Synchrotron (AGS) with a sample
of 10300 events and branching ratio (BR) of
$[2.94\pm0.05(stat)\pm0.13(syst)\pm0.05(theor)]\times 10^{-7}$.
For the muon channel, it was first observed by  BNL-E787 \cite{E787}
at AGS with the measured branching ratio being $(5.0\pm0.4\pm0.9)\times
10^{-8}$, which is too small to accommodate within the SM.
However,  two subsequent
experiments of BNL-E865 \cite{E865m}
and HyperCP (E871) \cite{E871}
have measured that $BR(K^+\to\pi^+\mu^+\mu^-)=(9.22\pm0.60\pm0.49)$
and $(9.8\pm1.0\pm0.5)\times 10^{-8}$, respectively,
which are all consistent with the model-independent
analysis based on the data of $BR(K^+\to\pi^+e^+e^-)$.
However, there are still rooms for new physics, particularly in
the muon mode.

In Refs. \cite{geng1,geng2}, $P$ and $T$ violating muon polarization
effects in $K^+\to \pi^+\mu^+\mu^-$
 were discussed in various theoretical models. In this paper,
we study the forward-backward asymmetry (FBA) in the decay
of $K^+\to \pi^+\ell^+\ell^-$ with $\ell=e$ or $\mu$.
It is known that the FBA in $K^+\to \pi^+\ell^+\ell^-$ violates $P$ like
the longitudinal lepton polarization but it vanishes in the SM and can
only exist if there is a scalar type interaction.
In the multi-Higgs doublet models such as the most popular two-Higgs
doublet (2HDM) of type II \cite{2HDM}, where two Higgs scalar doublets
($H_u$ and
$H_d$) are coupled to up- and down-type quarks, respectively, the scalar
type of four fermion operators
$\bar{s}_Rd_L\,\bar{\ell}\ell$ can be generated at the loop level
\cite{tanB0,tanB}.
 This type of operators is particularly interesting in the minimal
supersymmetric standard model (MSSM) \cite{MSSM}
since it receives an enormous
enhancement for the large ratio of $v_u/v_d =\tan\beta$ where $v_{u(d)}$
is the vacuum expectation value of the Higgs doublet $H_{u(d)}$.
Recently, there has been considerable interest for the large $\tan\beta$
effects in $B$ decays such as $B\to\mu^+\mu^-$
and $B\to K\mu^+\mu^-$ \cite{tanB0,tanB,Bmm}. In the report, we will
discuss the
large $\tan\beta$ scenario in the MSSM for $K^+\to\pi^+\ell^+\ell^-$.

The paper is organized as follows. In Sec.~II, we present the
general analysis for the forward-backward asymmetries in
$K^+\to\pi^+\ell^+
\ell^-$ ($\ell=e$ and $\mu$).
 In Sec.~III, we discuss the experimental constraints on
the asymmetries.
 We estimate the asymmetries in the MSSM in Sec.~IV.
We present our conclusions in Sec.~V.

\section{General analysis}
We write the decay as \be K^+(p_K)\to\pi^+(p_\pi)
\ell^+(p)\ell^-(\bar{p})\,, \ee where $p_K$, $p_\pi$, $p$ and
$\bar{p}$ are four-momenta of $K^+$, $\pi^+$, $\ell^+$ and
$\ell^-$, respectively. The most general invariant amplitude for
the decay can be written as \cite{geng1,geng2,geng3} \be {\cal
M}&=& F_{S}\bar{\ell}\ell+iF_{P}\bar{\ell}\gamma
_{5}\ell+F_{V}p_{K}^{\mu }\bar{\ell}\gamma
_{\mu}\ell+F_{A}p_{K}^{\mu }\bar{\ell}\gamma _{\mu }\gamma
_{5}\ell\,, \label{Am0} \ee where $F_S$, $F_P$, $F_V$ and $F_A$
are scalar, pseudo-scalar, vector and axial-vector form factors,
respectively. The differential decay rate in the $K^+$ rest frame
is given by \cite{geng1} \be \frac{d^{2}\Gamma }{dE\,d\bar{E}} &=&
\frac{1}{2^{4}\pi ^{3}m_{K}}\left[\left| F_{S}\right|
^{2}\frac{1}{2}(s-4m_l^{2})+\left| F_{P}\right| ^{2}\frac{1}{2}s
\right.
\nn\\
&&+\left| F_{V}\right| ^{2} m_{K}^{2} (2E \bar{E}-\frac{1}{2} s)
+\left| F_{A}\right|^{2} m_{K}^{2}(2 E \bar{E} - \frac{1}{2} s+
2m_l^{2})
\nn\\
&&\left. +2 \texttt{Re} (F_{S} F_{V}^{\ast })m_lm_{K}(E-\bar{E})
+\texttt{Im}(F_{P}F_{A}^{\ast }) m_l (m_{\pi }^{2}-m_{K}^{2}-s)
\right]\,, \label{rate1}
\ee%
where $m_l$ is the lepton mass, $E(\bar{E})$ is the energy of
$\mu^+(\mu^-)$ and $s=(p+\bar{p})^{2}=2(m_l^{2}+E\bar{E}-{\bf p}
\cdot \bar{{\bf p}})$ is the invariant mass of the dilepton
system. In terms of the invariant mass and the angle $\theta$
between the three-momentum of the kaon and the three-momentum of
the $\ell^-$ in the dilepton rest frame, we can rewrite Eq.
(\ref{rate1}) as \be \frac{d^{2}\Gamma }{ds\,d\cos\theta }
&=&\frac{1}{2^{8}\pi^{3}m_{K}^{3}}\cdot \beta_l \lambda^{1\over 2}
(s) \left\{\left|F_{S}\right|^{2}s\beta_l^{2}+\left|
F_{P}\right|^{2}s \right.
\nn\\
&&+\left|F_{V}\right|^{2}\frac{1}{4}\lambda (s)
(1-\beta_l^{2}\cos^2\theta )
+\left| F_{A}\right|^{2}\left[\frac{1}{4}\lambda (s)
(1-\beta_l^{2}\cos^2\theta )+4m_{K}^{2}m_l^{2}\right]
\nn\\
&& \left.+\texttt{Re}(F_{S}F_{V}^{\ast })2m_l\beta_l
 \lambda^{1\over 2} (s) \cos\theta
+\texttt{Im}(F_{P}F_{A}^{\ast })2m_l(m_{\pi
}^{2}-m_{K}^{2}-s)\right\}\,, \label{rate2} \ee where $\lambda
(s)=m_{K}^{4}+m_{\pi }^{4}+s^{2}-2m_{\pi
}^{2}s-2m_{K}^{2}s-2m_{\pi }^{2}m_{K}^{2}$ and
$\beta_l=(1-4m_l^2/s)^{1\over 2}$ with $s$ and $\cos\theta$
bounded by \be 4m_l^2\leq s\leq (m_K-m_\pi)^2\,,\ \
-1\leq\cos\theta\leq 1\,. \ee Here, we have used that \be
E=\frac{s+m_{K}^{2}-m_{\pi }^{2}+\beta_l \lambda^{1\over 2} (s)
\cos\theta}{4m_{K}%
},\qquad
\bar{E}=\frac{s+m_{K}^{2}-m_{\pi }^{2}-\beta_l \lambda^{1\over 2} (s)
\cos\theta}{4m_{K}}\,.
\ee
By integrating the angle $\theta$ in Eq. (\ref{rate2}), we obtain
\be
\frac{d\Gamma }{ds}
&=&\frac{1}{2^{8}\pi^{3}m_{K}^{3}}\cdot
\beta_l \lambda^{1\over 2} (s)
\left\{\left|F_{S}\right|^{2}2s\beta_l^{2}+\left| F_{p}\right|^{2}2s
+\left|F_{V}\right|^{2}\frac{1}{3}\lambda (s)
(1+{2m_l^2\over s})
\right.
\nn\\
&& \left. +\left| F_{A}\right|^{2}\left[\frac{1}{3}\lambda (s)
(1+{2m_l^2\over s}) +8m_{K}^{2}m_l^{2}\right]
+\texttt{Im}(F_{P}F_{A}^{\ast })4m_l(m_{\pi
}^{2}-m_{K}^{2}-s)\right\}\,. \label{rate3} \ee
 From
Eq. (\ref{rate2}) and the definition of the forward-backward asymmetry
\be
A_{FB}(s)&\equiv& \frac{\int_{0}^{1}d\cos\theta \frac{d^{2}\Gamma }
{dsd\cos\theta }%
-\int_{-1}^{0}d\cos\theta \frac{d^{2}\Gamma }{dsd\cos\theta
}}{\int_{0}^{1}d\cos\theta
\frac{d^{2}\Gamma }{dsd\cos\theta }+\int_{-1}^{0}d\cos\theta
\frac{d^{2}\Gamma }{%
dsd\cos\theta }}\,,
\ee
we find that
\be
A_{FB}(s)
&=&{1\over 2^8\pi^3m_K^3}\cdot 2m_l\beta_l^2 \lambda (s)
\texttt{Re}(F_SF_V^{\ast})
 \left({\frac{d\Gamma }{ds}}\right)^{-1}\,.
\label{As0}
\ee
As seen from Eq. (\ref{As0}), to get a nonzero value of $A_{FB}$, it is
necessary to have a scalar interaction. However, in the SM  the contributions
from $F_S$ to the decay widths of $K^+\to\pi^+ e^+e^-$ and
$K^+\to\pi^+ \mu^+\mu^-$ are about 7 and 4 orders of magnitude smaller
than those from $F_V$ \cite{ChPT2,Gao}, respectively, and therefore the
forward-backward asymmetries are expected to be vanishingly small.

\section{Experimental constraints}
To study the experimental constraints on $A_{FB}$ in
$K^+\to\pi^+\ell^+\ell^-$, we consider the amplitude
adopted in Ref. \cite{E865e};
\be
{\cal M}&=&\frac{\alpha G_{F}}{4\pi }f_{V}P^{\mu
}\bar{\ell}\gamma_{\mu
}\ell+G_{F}m_{K}f_{S}\bar{\ell}\ell+G_{F}f_{T}\frac{P^{\mu}
q^{\nu }}{m_{K}}\bar{\ell}\sigma _{\mu \nu }\ell
\label{Am1}
\ee
where $f_{V,S,T}$ are dimensionless form factors of
vector, scalar, and tensor interactions, respectively, $P=p_K+p_\pi$ and
$q=p_K-p_\pi$.
It is clear that, in Eq. (\ref{Am1}),
the vector term arises from the one photon exchange in the SM,
which gives the dominant contribution to the decay rate, whereas
the scalar and tensor ones from some new physics beyond the SM
\cite{Tensor}.

For the form factor $f_V$, we take the form derived in the ChPT
\cite{ChPT1}, given by
\be
f_V(s) &=& a_++b_+{s\over m_K^2}+\omega^{\pi\pi}(s)\,,
\ee
where $a_+$ and $b_+$ are free parameters and $\omega^{\pi\pi}$
is the contribution from a pion loop diagram given in Ref. \cite{ChPT1}.
The experimental measurement on $K^+\to\pi^+e^+e^-$
at BNL-E865 \cite{E865e} has determined the
parameters of $a_+$ and $b_+$ to be $-0.587\pm0.010$ and $-0.655\pm 0.044$,
respectively.
The scalar and tensor form factors in  Eq. (\ref{Am1})
for $K^+\to\pi^+e^+e^-$ are also constrained by
the experiment \cite{E865e} and the results are that
\be
|f_S|&<& 6.6\times 10^{-5}\,\ \ or\ \
|f_T|\;<\; 3.7\times 10^{-4}\,
\label{fsft}
\ee
for the existence of either scalar or tensor interaction.
We note that so far there are no similar
constraints on $f_{S,T}$ for $K^+\to\pi^+\mu^+\mu^-$
and they can be quite different for the two
 channels in theoretical models.

It is easy to see that the amplitude in Eq. (\ref{Am1}) can be simplified
to
\be
{\cal M}&=&\frac{\alpha G_{F}}{4\pi }f_{V}^{\prime }P^{\mu }%
\bar{\ell}\gamma _{\mu }\ell+G_{F}m_{K}f_{s}^{\prime
}\bar{\ell}\ell
\label{Am2}
\ee
with
\be
f_{V}^{\prime }&=&f_{V}-\frac{8\pi im_l }{\alpha m_{K}}f_T,\qquad
f_{S}^{\prime }=f_{S}-
{i\beta_l\lambda^{1\over 2}(s)\cos\theta\over m_K^2}f_T\,.
\ee
By comparing the amplitude in Eq. (\ref{Am2}) with the general one
in Eq. (\ref{Am0}), we get
\be
F_{V}=
\frac{%
\alpha G_{F}}{2\pi }f_{V}^{\prime }\,,\ \
F_{S}= G_{F}m_{K}f_{S}^{\prime }\,,\ \ F_{P,A}=0\,.
\label{FF1}
\ee
From Eqs. (\ref{rate2}), (\ref{rate3}) and (\ref{FF1}), we obtain
\be
{d^2\Gamma \over ds\,d\cos\theta}
&=& {G_F^2\over 2^8\pi^3m_K^3}\cdot
\beta_l\lambda^{1\over 2} (s)
\left\{\left|f_V\right|^2{\alpha^2\over 16\pi^2}
\lambda (s)(1-\beta_l^2\cos^2\theta)+\left|f_S\right|^2s\beta_l^2m_K^2
\right.
\nn\\
&& +|f_T|^2 {s\lambda (s)\over m_K^2} (\cos^2\theta+{4m_l^2\over
s}\sin^2\theta) +\texttt{Re}(f_V^{\ast}f_S){\alpha m_lm_K\over
\pi}\beta_l \lambda^{1\over 2} (s)\cos\theta
\nn\\
&& \left. -\texttt{Im}(f_Vf_T^{\ast}) {\alpha\lambda (s)\over
\pi}{m_l\over m_K}-\texttt{Im}(f_Sf_T^{\ast})
2s\beta\lambda^{1\over 2} (s)\cos\theta \right\} \, \ee and \be
\frac{d\Gamma }{ds} &=&\frac{G_{F}^{2}}{2^{8}\pi
^{3}m_{K}^{3}}\cdot \beta_l \lambda^{1\over 2} (s) \left\{\left|
f_{V}\right| ^{2}\frac{\alpha ^{2}\lambda (s)
}{%
4\pi ^{2}}\frac{1}{3}(1+\frac{2m_l^{2}}{s})+2\left| f_{S}\right|
^{2}s\beta_l
^{2}m_{K}^{2}\right.
\nn\\
&&
\left.+\left| f_{T}\right| ^{2} {2s\lambda (s)\over 3m_{K}^{2}}
(1+{8m_l^2\over s})
-\texttt{Im}(f_{V}f_{T}^{\ast }) \frac{2\alpha \lambda (s) }{%
\pi }\frac{m_l}{m_{K}} \right\}\,. \label{ratef} \ee Similarly,
from Eq. (\ref{As0}) we find \be A_{FB}(s)&=&
\frac{G_{F}^{2}}{2^{8}\pi ^{3}m_{K}^{3}}\cdot \beta_l^2 \lambda
(s) \left[ \texttt{Re}(f_{V}^{\ast }f_{S})\frac{\alpha
m_lm_{K}}{\pi }%
-\texttt{Im}(f_{S}f_{T}^{\ast }) 2s
 \right] \left({\frac{d\Gamma }{ds}}\right)^{-1}\,.
\ee
 From Eq. (\ref{ratef}),
one can check that the bound for $f_S$ or $f_T$ in Eq.
(\ref{fsft}) yields at most a few percent of the decay rate in
$K^+\to\pi^+e^+e^-$. Moreover, the last term in Eq.
(\ref{ratef}) is negligible for the electron channel no matter
whether $f_T$ is real or imaginary due to the electron mass suppression.
However, for the muon case this term could be large
 and spoil the vector dominant mechanism if the imaginary part of
$f_T$ is not small.
In Figure 1, we show the differential decay rate and forward-backward
asymmetry as functions
of $\hat{s}=s/m_K^2$ for the decay of $K^+\to\pi^+e^+e^-$
by using the upper value of $f_S$ in
Eq. (\ref{fsft}) and $f_T=0$. In Figure 2, we display them
 by assuming that
$f_{s}\sim -4\times 10^{-5}i$ and
$f_{T}\sim 2\times 10^{-4}$.
As illustrations, in Figures 3 and 4 we also give
$d\Gamma/d\hat{s}$
and $A_{FB}$ in $K^+\to\pi^+\mu^+\mu^-$ with the same sets
of parameters as those in Figures 1 and 2, respectively.
 It is clear that,
as mentioned early, since there is no direct strict experimental
constraint on
$f_S$ or $f_T$ in the muon mode, $A_{FB}(K^+\to\pi^+\mu^+\mu^-)$ can be
arbitrary large.

\section{Supersymmetry}

In the MSSM, the one-loop effective down-type Yukawa interaction
is given by
\be
{\cal L}^{eff}&=& \bar{d}_RY_d[H_d+(\epsilon_0+\epsilon_YY_u^{\dagger}Y_u)
H^*_u]Q_L\;+\;h.c.,
\label{L1}
\ee
where $Y_{u,d}$ are $3\times 3$ Yukawa coupling matrices and
$\epsilon_{0,Y}$ are defined in Ref. \cite{tanB},
which are typically ${\cal O}(10^{-2})$. In the diagonal $Y_d$ basis
of $(Y_d)_{ij}=y^d_i\delta_{ij}$,
the interaction in Eq. (\ref{L1}) becomes
\be
{\cal L}^{eff}_{mass}&=&
v_d
\bar{d}^i_Ry_i^d\left[(1+\epsilon_0\tan\beta)\delta_{ij}+
\epsilon_Y V_{ik}^{\dagger}(y_k^u)^2V_{kj}\right]d_L^j
\;+\;h.c.,
\label{L2}
\ee
where $V$ is the CKM mixing matrix.
By writing the effective Hamiltonian in the transition $s\to
d\ell^+\ell^-$ induced by the scalar type of interactions as
\be
{\cal H}^{eff}_{S}&=&
(C_S\bar{s}_Rd_L+C_S'\bar{d}_Rs_L)\,\bar{\ell}\ell\,,
\label{Ham}
\ee
 from Eq. (\ref{L2}) one has that \cite{tanB}
\be
C_S&=& -{G_F^2\over 4\pi^2}
{m_sm_lm_t^2\bar{\lambda}^t_{21}\tan^3\beta
\over (1+\epsilon_0\tan\beta)} \,{1\over M_A^2}\,{\mu\, A\, f(x_{\mu
L},x_{\mu R})\over M^2_{\tilde{t}_L}}\,,
\nn\\
C_S' &\simeq & {m_d\over m_s}C_S\,,
\ee
where
 $A$ is the coupling of the soft-breaking trilinear term and
\be
\bar{\lambda}^t_{21}&=& \lambda^t_{21}
\left[{1+\tan\beta (\epsilon_0+
\epsilon_Yy_t^2)\over 1+\epsilon_0\tan\beta}\right]^2
\nn\\
f(x,y)&=& {1\over x-y}\left[{x\ln x\over 1-x}-{y\ln y\over 1-y}\right]\,,\
f(1,1)\;=\;{1\over 2}\,,
\ee
with
\be
x_{{\mu L}}&=& {\mu^2\over M^2_{\tilde{t}_L}}\,,\ \
x_{{\mu R}}\;=\;{M^2_{\tilde{t}_R}\over M^2_{\tilde{t}_L}}\,,\ \
\lambda^t_{21}\;=\; V_{ts}^*V_{td}\,,
\ee
and $y_t$ being the top quark Yukawa coupling.
By comparing Eq. (\ref{Ham}) with Eq. (\ref{Am1}) and using
\be
<\pi|\bar{d}(1+\gamma_5)s|K>&\simeq & {m_K^2\over m_s}f_+\,,
\ee
we find
\be
f_S^{MSSM}&\simeq&
 -{G_F^2\over 8\pi^2}
{m_Km_lm_t^2\bar{\lambda}^t_{21}\tan^3\beta
\over (1+\epsilon_0\tan\beta)^2} \,{1\over M_A^2}\,{\mu\, A\, f(x_{\mu
L},x_{\mu R})\over M^2_{\tilde{t}_L}}\,,
\label{fsMSSM}
\ee
where we have neglected the small terms related to $y_{1,2}^u$ and
used $f_+\simeq 1$.

To estimate  the scalar form factor
in Eq. (\ref{fsMSSM}) in the MSSM with large $\tan\beta$, we take
$\epsilon_0\sim 1/100\gg \epsilon_Yy_t^2$
and $\tan\beta=50r$ and we get
\be
f_S^{MSSM}|_{\ell=e}&\sim&
 1.1\times 10^{-9}(1-\bar{\rho}-i\bar{\eta})
{r^3\over (1+{1\over 2}r)^2}
\,\left({200\ GeV\over M_A}\right)^2\,\left({\mu\, A\, f(x_{\mu
L},x_{\mu R})\over M^2_{\tilde{t}_L}}\right)\,,
\nn\\
f_S^{MSSM}|_{\ell=\mu}&\sim&
 2.3\times 10^{-7}(1-\bar{\rho}-i\bar{\eta}){r^3\over (1+{1\over 2}r)^2}
\,\left({200\ GeV\over M_A}\right)^2\,\left({\mu\, A\, f(x_{\mu
L},x_{\mu R})\over M^2_{\tilde{t}_L}}\right)\,,
\label{fsMSSMn}
\ee
where $\bar{\rho}=\rho (1-\lambda^2/2)$ and $\bar{\eta}=\eta
(1-\lambda^2/2)$ with $\lambda$, $\rho$ and $\eta$ being
the Wolfenstein parameters of the CKM matrix $V$.
Since the values of $f_S^{MSSM}|_{\ell=e,\mu}$ in Eq. (\ref{fsMSSMn}) are
about three and one
orders of magnitude smaller than the experimental bound in  Eq.
(\ref{fsft}) and thus the scalar contributions to the decay rates
in the MSSM are negligible, respectively. Moreover,
the scalar contribution to the FBA
in $K^+\to\pi^+e^+e^-$ is also suppressed. However, in
$K^+\to\pi^+\mu^+\mu^-$
the FBA can be as large as $
10^{-3}$ as shown in
Figure 5
by using $\bar{\rho}
\sim 0.2$ \cite{Buras02} and assuming $r\sim 1$,
$M_A\sim 200\
GeV$ and
$\mu A f(x_{\mu L},x_{\mu R})/ M^2_{\tilde{t}_L}\sim 2$.
We note that $A_{FB}(K^+\to\pi^+\mu^+\mu^-)={\cal O}(10^{-3})$ is
accessible to future experiments such as the CKM at Fermilab, where
the order of $10^5$ events can be produced \cite{Kaon}.

\section{Conclusions}
We have studied the forward-backward asymmetries in the decays of
$K^+\to\pi^+\ell^+ \ell^-$ ($\ell=e$ and $\mu$) in the most general
amplitudes. In particular, we have explored the experimental constraints
on the asymmetries by including the scalar and tensor interactions.
We have found that with the scalar (tensor) term
the asymmetry can be up to ${\cal O}(10^{-3})$ (${\cal O}(10^{-1}))$
and arbitrary large for the electron and muon channels, respectively,
without conflict with the experimental data.
We have also discussed the asymmetries in the minimal supersymmetric
standard model where the scalar terms can be explicitly induced.
We have shown that the FBA in $K^+\to\pi^+e^+ e^-$
is negligibly small due to the electron mass suppression, but in
$K^+\to\pi^+\mu^+ \mu^-$ it
can be as large as ${\cal O}(10^{-3})$ with the large
$\tan\beta$,
 which can be tested in future experiments such as
the CKM experiment at Fermilab.\\



 \noindent {\bf Acknowledgments}

We thank Drs. D.N. Gao and H.K. Park for useful communications
and Prof. T. Han for encouragement and discussions.
This work was supported in part by
 the National Science Council of the Republic of China under
 Contract Nos. NSC-91-2112-M-001-053 and NSC-91-2112-M-007-043.


\newpage

\begin{figure}[tbp]
\vspace{0cm} \centerline{ \psfig{figure=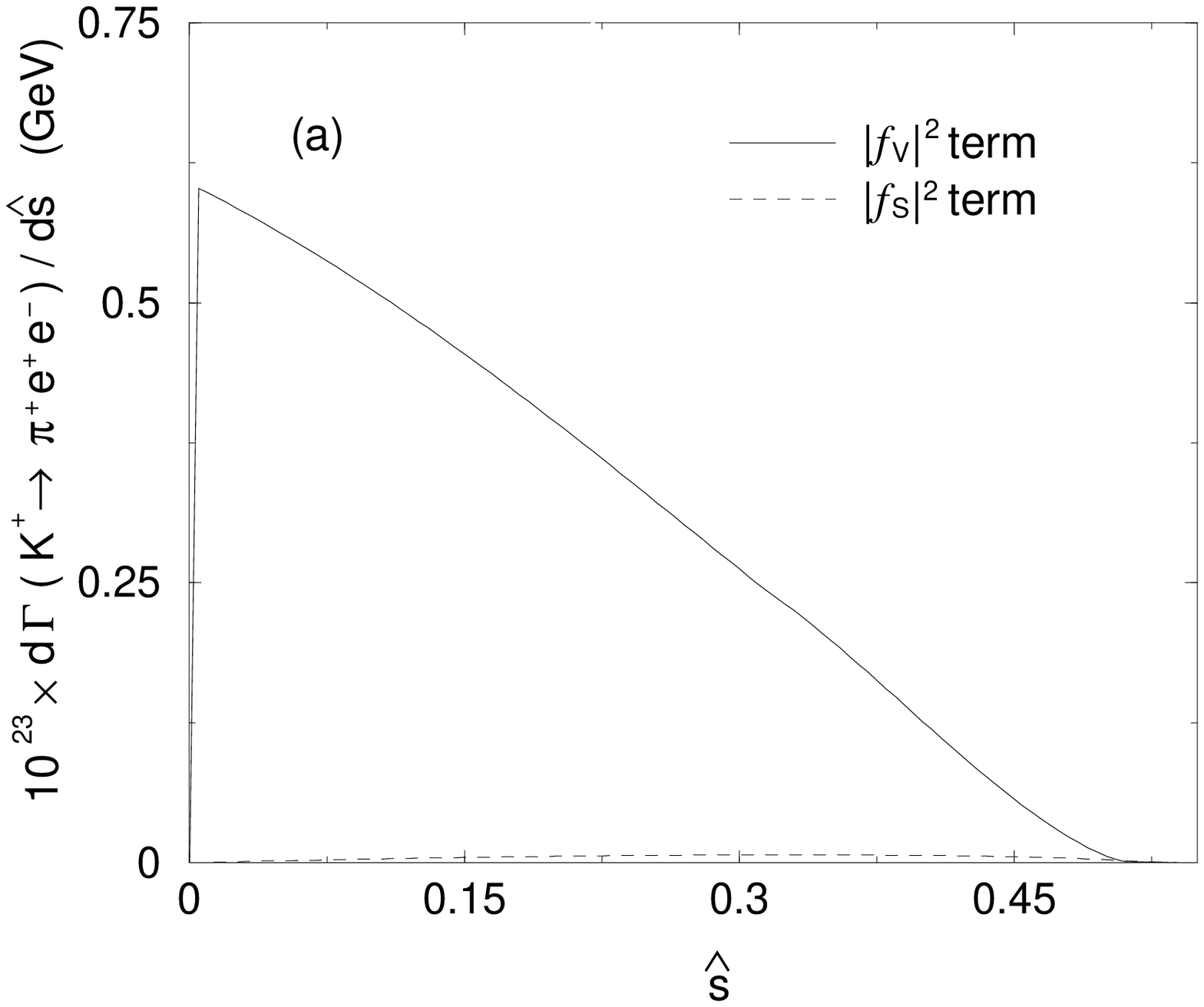,height=2.6in }
$\ \ $ \psfig{figure=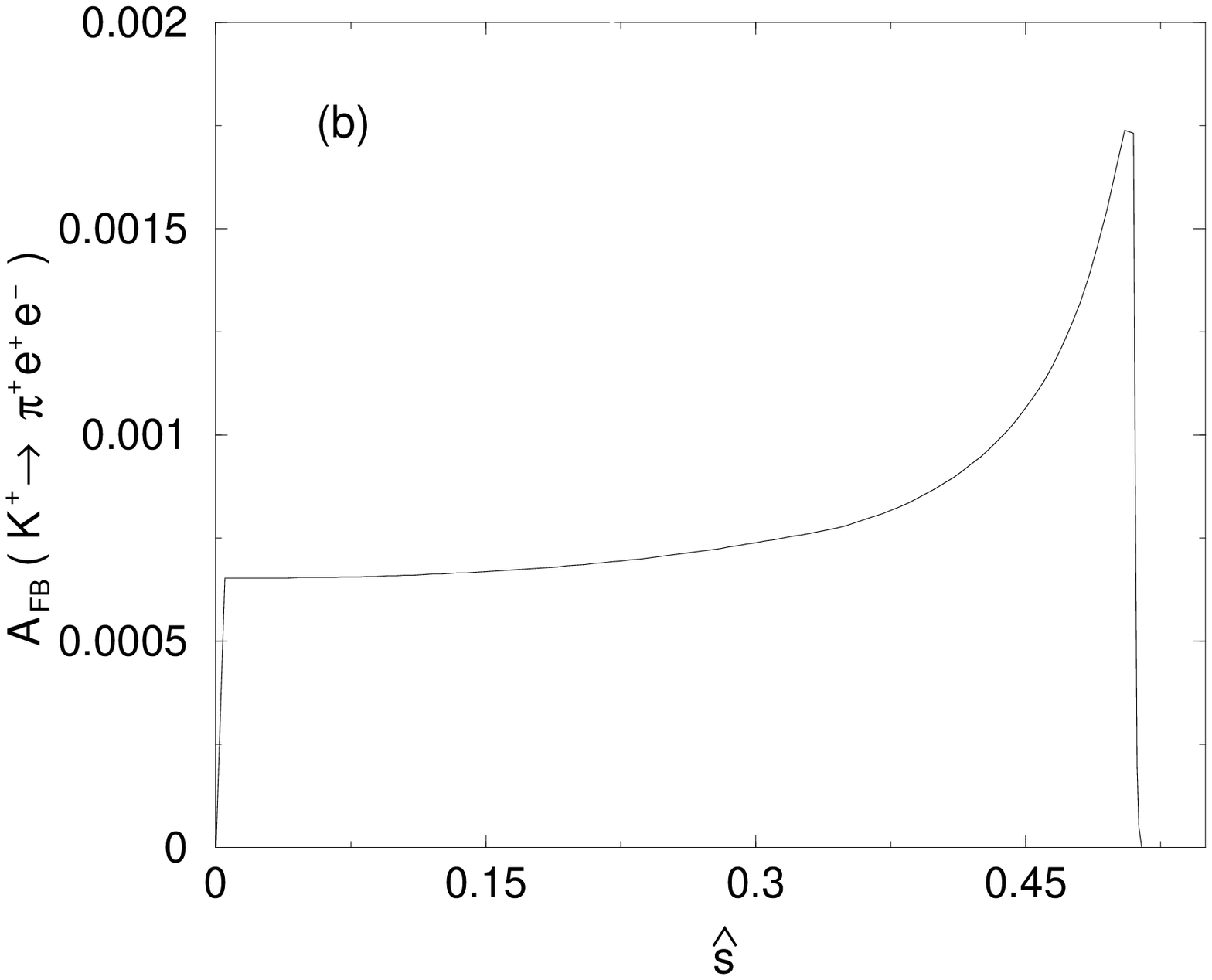,height=2.6in } } \caption{(a)
Differential decay rate and (b) forward-backward asymmetry for
$K^+\to \pi^+e^+e^-$ as functions of $\hat{s}=s/m_K^2$ with
$f_S=6.6\times 10^{-5}$ and $f_T=0$.}
\label{Figure1}
\end{figure}
\begin{figure}[tbp]
\vspace{3cm} \centerline{ \psfig{figure=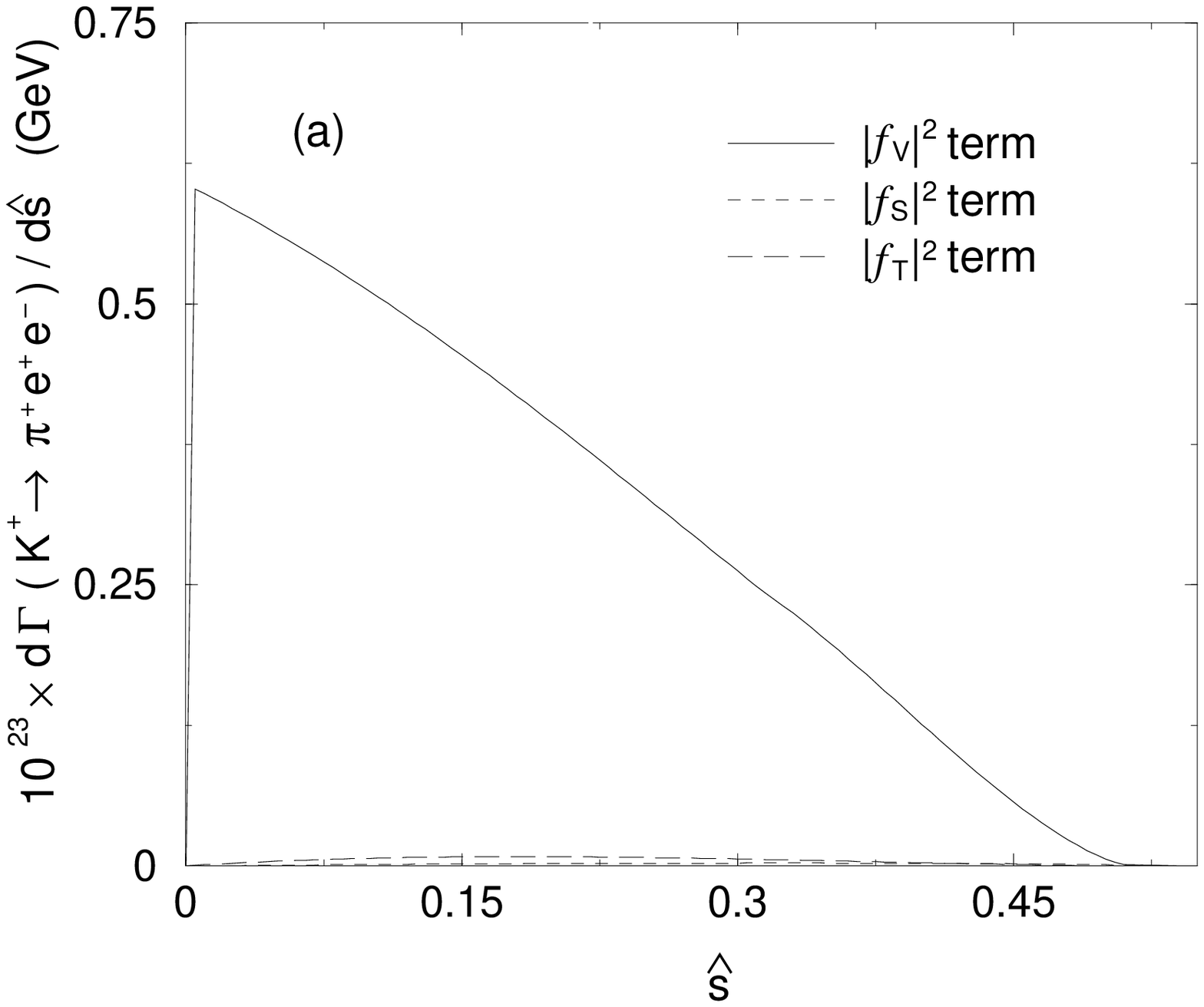,height=2.6in }
$\ \ $ \psfig{figure=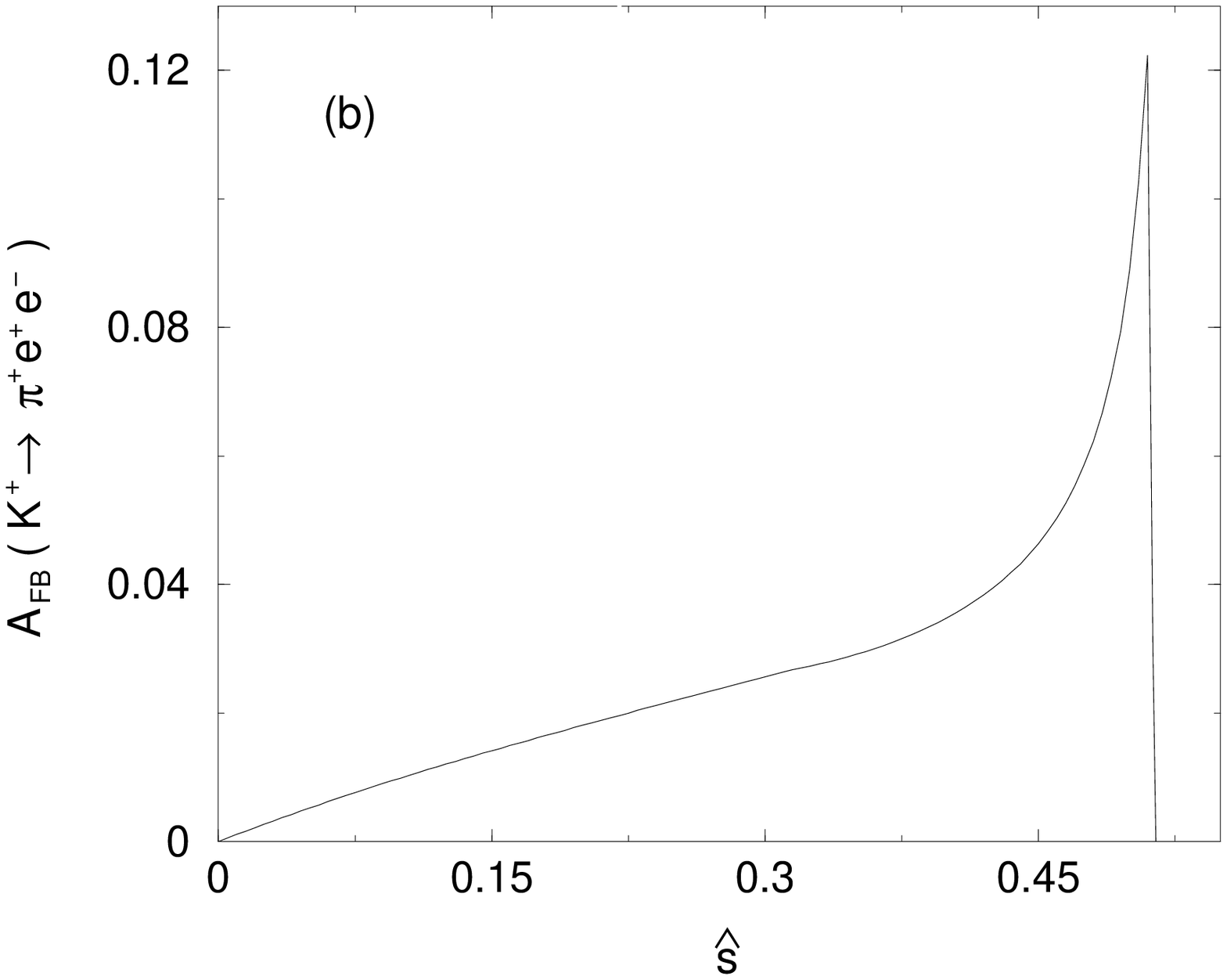,height=2.6in } } \caption{Same as
Figure \ref{Figure1} but $f_{S}\sim -4\times 10^{-5}i$ and
$f_{T}\sim 2\times 10^{-4}$} \label{Figure2}
\end{figure}

\newpage

\begin{figure}[tbp]
\vspace{0cm} \centerline{ \psfig{figure=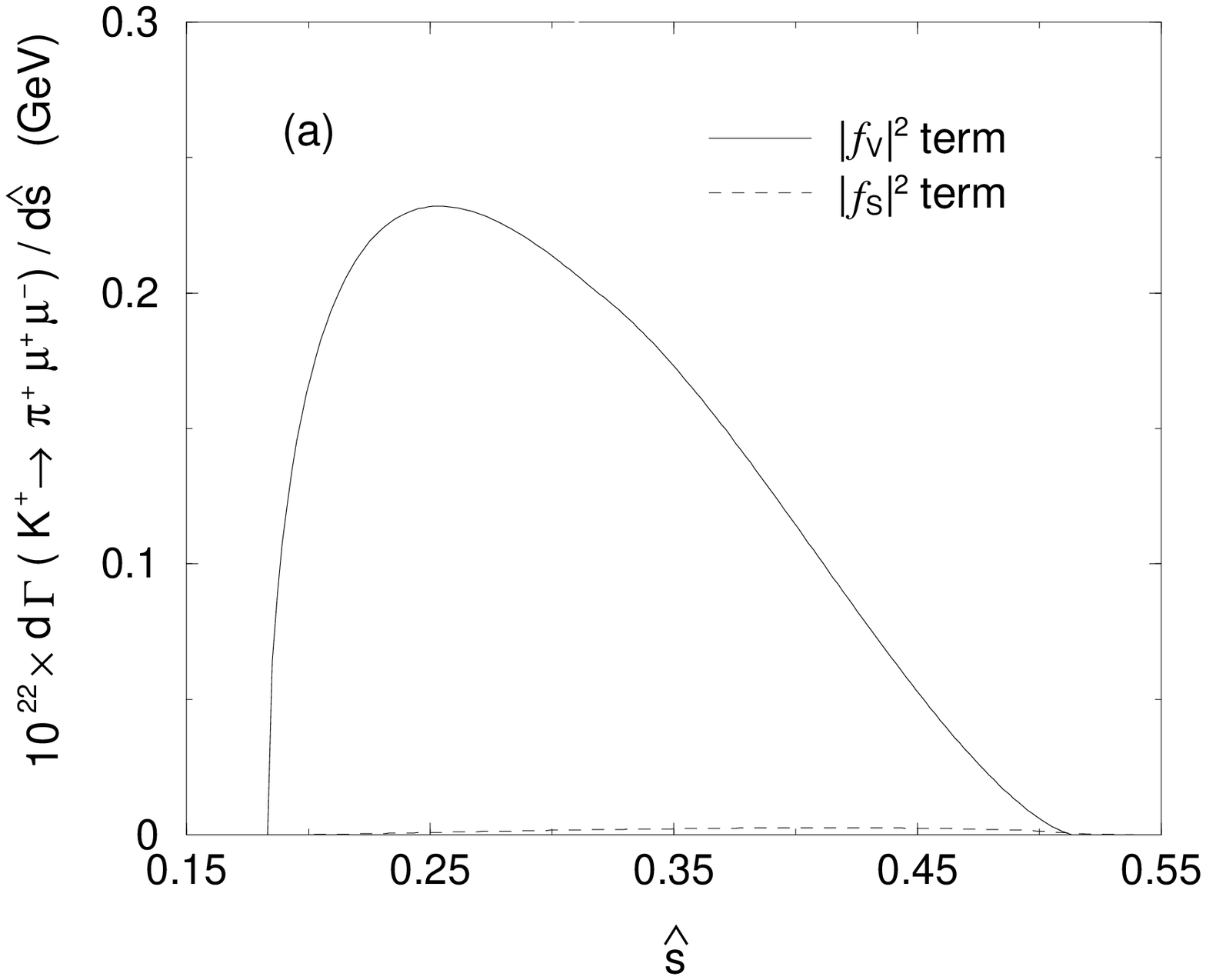,height=2.6in }
$\ \ $ \psfig{figure=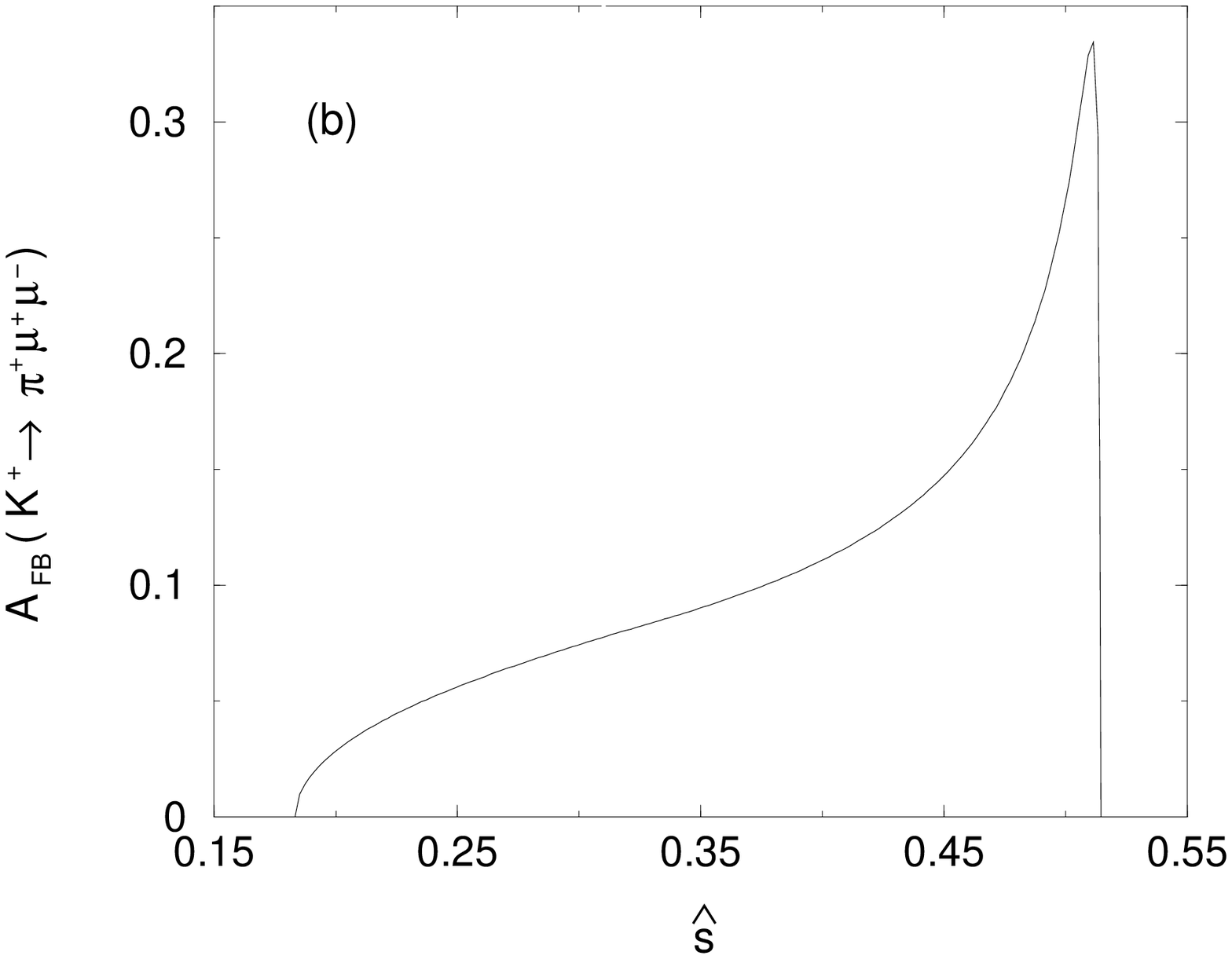,height=2.6in } } \caption{Same as
Figure 1 but for
 $K^+\to\pi^+ \mu^+ \mu^-$.}
\label{Figure3}
\end{figure}

\begin{figure}[tbp]
\vspace{0cm} \centerline{ \psfig{figure=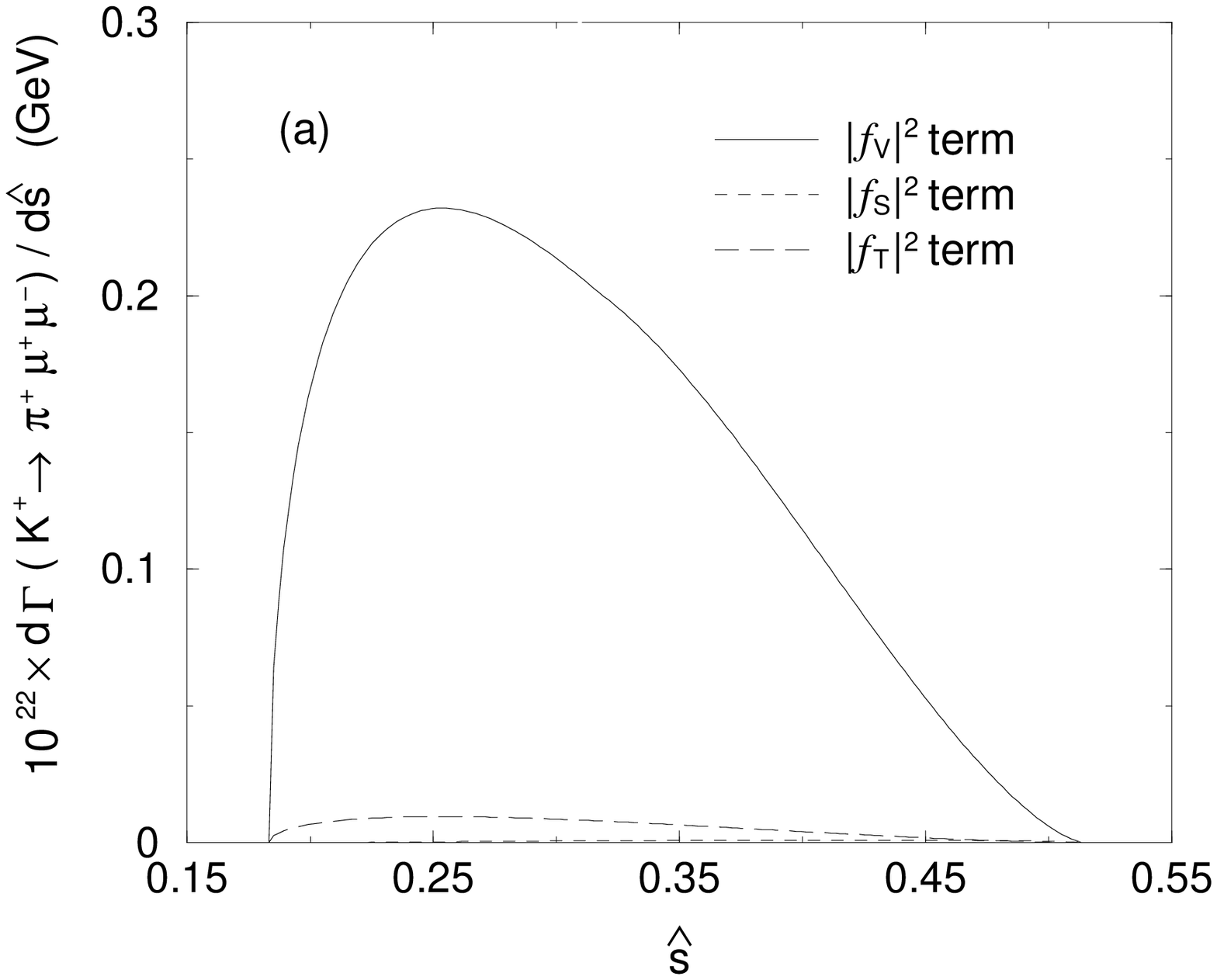,height=2.6in }
$\ \ $ \psfig{figure=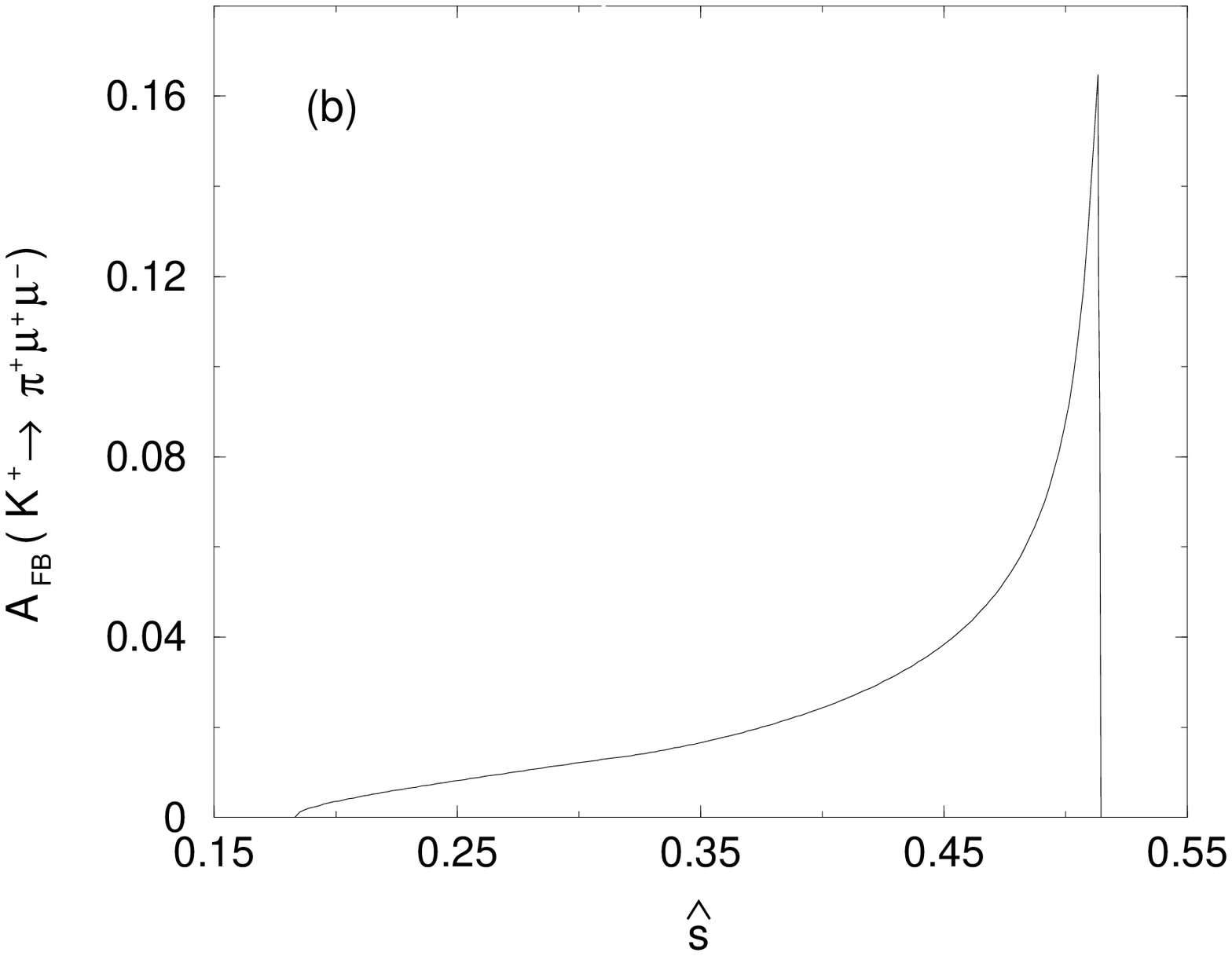,height=2.6in } } \caption{ Same as
Figure 2 but for
 $K^+\to\pi^+ \mu^+ \mu^-$.}
\label{Figure4}
\end{figure}

\begin{figure}[tbp]
\vspace{2cm} \centerline{ \psfig{figure=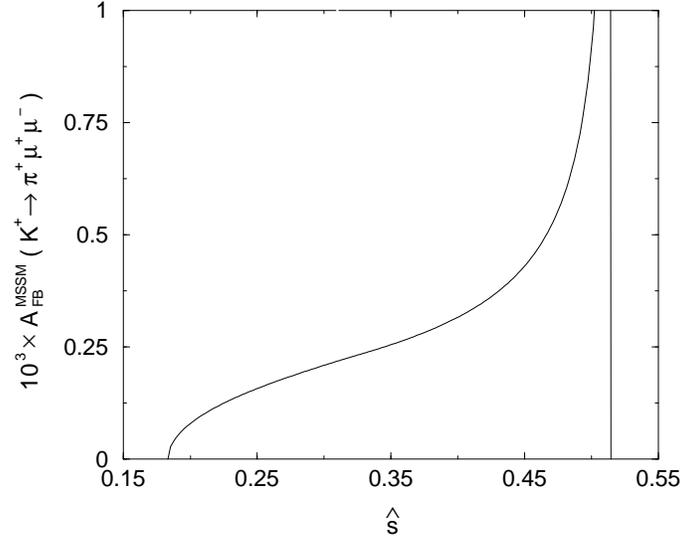,height=3.in } }
\caption{
Forward-backward asymmetry
in $K^+\to \pi^+\mu^+\mu^-$ as a function of $\hat{s}$
in the MSSM with large $\tan\beta$.
}
\label{Figure5}
\end{figure}

\end{document}